\pdfoutput=1
\documentclass[aps,prd,12pt,nofootinbib]{revtex4}
\usepackage{epsfig}
\usepackage{graphicx}
\usepackage{amsmath}
\usepackage{amssymb}
\usepackage{mathrsfs}
\usepackage{verbatim}
\usepackage{float}
\usepackage{comment}

\usepackage[normalem]{ulem}
\usepackage{xcolor}

\newcounter{fig}

\newcommand{\bea}{\begin{eqnarray}}
\newcommand{\eea}{\end{eqnarray}}
\newcommand{\be}{\begin{equation}}
\newcommand{\ee}{\end{equation}}

\def\pa{\partial}

\def\({\left(}
\def\){\right)}

\newcommand{\re}[1]{(\ref{#1})}

\newcommand{\tr}{\mbox{Tr}}

\def\rlx{\relax\leavevmode}
\def\IR{\rlx\hbox{\rm I\kern-.18em R}}
\def\one{\hbox{{1}\kern-.25em\hbox{l}}}

\newcommand{\eqn}{\begin{eqnarray}}
\newcommand{\eqnx}{\end{eqnarray}}

\date{\today}

\begin{document}
\title{Multisolitons  in a gauged Skyrme-Maxwell model}
\author{L. R. Livramento}
\affiliation{BLTP, JINR, Dubna 141980, Moscow Region, Russia}
%\author{E.~Radu}
%\affiliation{Department of Mathematics, University of Aveiro and CIDMA,
%Campus de Santiago, 3810-183 Aveiro, Portugal}
\author{Ya.~Shnir}
\affiliation{BLTP, JINR, Dubna 141980, Moscow Region, Russia\\
%Department of Theoretical Physics, Tomsk State %Pedagogical University, Russia
Institute of Physics,
Carl von Ossietzky University Oldenburg, Germany
Oldenburg D-26111, Germany}

\begin{abstract}
We construct new solutions of a $U(1)$ gauged Skyrme-Maxwell model of topological degrees $Q\le 5$, which represent Skyrmions
coupled to magnetic fluxes.
It is found that, generically, in the strong coupling limit, coupling to the magnetic field results in
 transformation of the configuration to the toroidally shaped
Skyrmions bounded to the local magnetic flux.
\end{abstract}
\maketitle
%%%%%%%%%%%%%%%%%%%%%%%%%%%%%%%%%%%%%%%%%%%%%%%%%%%%%%%%%%
\section{Introduction}
%%%%%%%%%%%%%%%%%%%%%%%%%%%%%%%%%%%%%%%%%%%%%%%%%%%%%%%%%%
The celebrated Skyrme  model \cite{skyrme,Skyrme:1962vh}  serves
as a prototype of a (3+1) dimensional non-linear field theory
supporting topological solitons.  This model possesses a wide
variety of applications in several areas of physics (for a review,
see \cite{Manton:2004tk,Manton:2022,Brown:2010api,Shnir:2018yzp}).
It has been suggested \cite{Witten:1983tw,Witten:1983tx} that the
model may be of relevance for a low energy description of QCD, as
the solitons are interpreted as nucleons and nuclei with the identification
of the topological charge with the baryon number.

In its simplest form, the Skyrme model contains only three
free parameters which set the length and
energy scales and the mass of the scalar excitations, respectively.
It successfully reproduces a number of qualitative features of nuclei
with surprising accord with experiment \cite{Brown:2010api,Zahed:1986qz}.
However, the standard version of the Skyrme model is plagued with problems of
prediction of binding energies for baryons and formation of cluster structures of nuclei.

Various modifications of the Skyrme model were considered over last two decades to improve the situation.
Among them are modifications of the potential
\cite{Battye:2006tb,Battye:2006na,Gudnason:2016yix,Gudnason:2015nxa,Dupuis:2018utr,Livramento:2022zly,Gudnason:2016mms,Gudnason:2016cdo},
extended theories which incorporate scalar and vector mesons
\cite{Adkins:1983nw,Jackson:1985yz,Schwesinger:1986xv,Yabu:1989un,Forkel:1991mv,Gudnason:2020arj,Naya:2018mpt,Naya:2018kyi}
and holographic models \cite{Sutcliffe:2010et,Sutcliffe:2015sta,Sutcliffe:2011ig}
inspired by the Atiyah-Manton construction of Skyrmions \cite{Atiyah:1989dq,Sakai:2004cn}.
It has been also suggested to include in the Lagrangian of the theory
higher-order terms with additional couplings
\cite{Marleau:1991jk,Marleau:1990nh,Floratos:2001ih,Adam:2010fg,Adam:2013wya,Adam:2013tga,Gudnason:2017opo}
or modify the Skyrme model to the form
which supports self-duality equations
\cite{Ferreira:2017yzy,Ferreira:2017bsr,Ferreira:2021ryf}.

The usual pion mass potential of the Skyrme model breaks the symmetry to the $SU(2)$ subgroup. This global symmetry can be gauged, inclusion of the Maxwell term leads to the
$U(1)$ gauged Skyrme model which has been proposed in \cite{Piette:1997ny,Radu:2005jp}.
This approach is similar to the construction of the
topological solitons  in the $U(1)$ gauged $O(3)$ sigma model  in (2+1) dimensions \cite{Schroers:1995he,Tchrakian:1995np,Schroers:1996zy},
in the planar Skyrme-Maxwell model \cite{Gladikowski:1995sc,Shnir:2015twa,Samoilenka:2015bsf} and its modifications
\cite{Adam:2012pm,Samoilenka:2016wys,Navarro-Lerida:2018siw,Navarro-Lerida:2018giv,Adam:2017ahh}.
Static soliton solutions of the Faddeev-Skyrme-Maxwell theory in 3+1 dimensions were discussed in
\cite{Shnir:2014mfa,Samoilenka:2018oil}.
In the context of the Skyrme model such a modification was originally motivated by construction of a
semiclassical model of interaction between a monopole and nucleons, so-called
Callan-Rubakov effect of the baryon decay catalysis
\cite{Rubakov:1981rg,Callan:1982ah}.
Notably, extended Skyrme-Maxwell theory
also can be derived in a holographic model via an expansion of a
Yang-Mills field of calorons \cite{Cork:2021ylu}.

In \cite{Piette:1997ny,Livramento:2023keg} it was shown that
coupling to the electromagnetic field  significantly affects properties of the Skyrme
field, even in the sector of topological degree one. In particular, it
violates the spherical symmetry of the configuration of topological degree one and induces a  magnetic moment of the Skyrmion.
Gauged Skyrmion possesses an electric charge and carries a magnetic flux.
The existence of such solutions relies on the presence of the
potential, the electrostatic potential
is restricted from above by the value of the pion mass \cite{Livramento:2023keg}. However, electrically neutral solutions
coupled to a magnetic flux, exist in the limit of zero potential for arbitrary values of the gauge coupling.

The purpose of the current paper is to extend this analysis by analyzing
how coupling to the magnetic flux affects the geometrical shape and energy
of the static multisoliton solutions of the $U(1)$ gauged Skyrme model
of topological degrees $Q=1-5$.

This paper is organised as follows. In section \ref{sec:model} we introduce the
model and define the topological charge of the gauged Skyrmions.  In section \ref{sec:3} we
define the parametrization of the fields
to be used in finding the numerical solutions and introduce reduced axially-symmetric Ansatz. In section \ref{sec:4} we present our full 3d numerical scheme and
discuss the numerical results. Finally, in section \ref{sec:5} we present
our conclusions and further remarks.

\section{The model}
\label{sec:model}

The Lagrangian density of the (3 + 1)-dimensional, $U(1)$ gauged massive Skyrme-Maxwell model
is defined as \cite{Callan:1983nx,Piette:1997ny,Radu:2005jp,Livramento:2023keg}
\be
\begin{split}
\label{Lag0}
{\cal L}&=- \frac{1}{4}\,{\cal F}_{\mu\nu}\,{\cal F}^{\mu\nu}+ \frac{f_\pi^2}{2}\,{\cal D}_\mu \phi^a {\cal D}^\mu \phi^a
-\frac{1}{4\,a_0^2}\,\left[ ({\cal D}_\mu \phi^a {\cal D}^\mu \phi^a)^2 -
 ({\cal D}_\mu \phi^a {\cal D}_\nu \phi^a)({\cal D}^\mu \phi^b {\cal D}^\nu \phi^b)\right]\\
&- m_\pi^2\,(1-\phi_0) \, ,
\end{split}~
\ee
Here we employ the $O(4)$ sigma-model representation  of the Skyrme field as an $S^3$ valued field
$\phi^a = (\phi_0,\phi_k)$ subject to the constraint $\phi^a\cdot \phi^a = 1 $. In the context of
application of the Skyrme model to nuclear physics, the component $\phi_0$
has the interpretation as a scalar meson while the isotriplet $\phi_k,$ $k=1,\,2,\,3$ corresponds to
the pions.
Note that the inclusion of the potential term is necessary to stabilize the model both with respect to
isorotations \cite{Battye:2005nx,Battye:2014qva} and coupling to the electromagnetic field \cite{Livramento:2023keg}.

The electromagnetic field strength tensor is
${\cal F}_{\mu\nu}=\pa_\mu {\cal A}_\nu-\pa_\nu {\cal A}_\mu$, $a=0,...,\,3$ and
the covariant derivative of scalar field $\phi^a$
is defined in terms of the four-potential ${\cal A}_\mu$ as \cite{Piette:1997ny,Radu:2005jp}
\be
{\cal D}_\mu \phi_\alpha = \pa_\mu \phi_\alpha -e\,{\cal A}_\mu\,\varepsilon_{\alpha\beta}\,\phi_\beta \,,
\qquad {\cal D}_\mu \phi_A = \pa_\mu \phi_A, \qquad \alpha,\,\beta=1,\,2,\,A=0,\,3
\label{covariant0}~.
\ee
The physical vacuum corresponds to ${\cal D}_\mu \phi_\alpha=0$, ${\cal F}_{\mu\nu}=0$ and $\phi_0=1, ~ \phi_k=0$.

In natural units the Skyrme constant $a_0$ is dimensionless, while the pion decay constant $f_\pi$
and the pion mass parameter $m_\pi$ are of dimension of mass. Two of these parameters may be scaled away
by introducing the energy and length scales $f_\pi/(4\,a_0)$ and $2/(a_0\,f_\pi)$, respectively.
The rescaled pion mass parameter is
$m=2\,m_\pi/(a_0\,f_\pi)$ and the gauge coupling becomes $g\equiv e/a_0$.
Then the rescaled Lagrangian of the Skyrme-Maxwell model can  be written as
\be
\begin{split}
\label{Lag}
{\cal L}&=- \frac{1}{2}\,F_{\mu\nu}\,F^{\mu\nu}+ D_\mu \phi^a D^\mu \phi^a
-\frac12 (D_\mu \phi^a D^\mu \phi^a)^2 +
\frac{1}{2} (D_\mu \phi^a D_\nu \phi^a)(D^\mu \phi^b D^\nu \phi^b)\\
&- 2\,m^2\,(1-\phi_0) \, ,
\end{split}~
\ee
where
\be
D_\mu \phi_\alpha = \pa_\mu \phi_\alpha -g\,A_\mu\,\varepsilon_{\alpha\beta}\,\phi_\beta \,,
\qquad D_\mu \phi_A = \pa_\mu \phi_A, \qquad \alpha,\,\beta=1,\,2,\,A=0,\,3
\label{covariant}~.
\ee
The Skyrme-Maxwell model \re{Lag} is invariant with respect to the local $U(1)$ gauge transformations
\be
\phi_1+i\,\phi_2\rightarrow e^{-i\,g\,\alpha}\,\(\phi_1+i\,\phi_2\),
\qquad A_\mu \rightarrow A_\mu + \pa_\mu \alpha
\label{gauge}~,
\ee
where $\alpha$ is any real function of coordinates. We will fix the gauge  setting
$A_0(\infty)= 0$.

The  model \re{Lag} has a conserved, integer-valued topological charge, the degree of the map
$S^3 \mapsto S^3$ which can be written as
\be
Q_T= \int d^3x \, q(\vec{r})= -\frac{1}{12\,\pi^2}\,
\int d^3x \,\varepsilon_{abcd}\,\varepsilon_{ijk}\,\phi^a\,\pa_i \phi^b\,\pa_j \phi^c\,\pa_k \phi^d\,  .
\label{chargemain0}
\ee
However, such a quantity is not invariant with respect the $U(1)$ gauge transformations \eqref{gauge}. 
Let us introduce the quantity
\be
Q_g=-\frac{1}{24\,\pi^2}\,\int d^3x \,\varepsilon_{ijk}\,\tr \(D_i U \,U^{-1}\,D_j U \,U^{-1}\,D_k U \,U^{-1}\) \,,
\label{Qg} \ee
where $U=\phi_0\,\one + i\,\phi_i\,\tau_i$ is the Skyrme field, $D_iU = \pa_iU -i\,g\,A_i\,\left[Q_c,\, U\right]$,  $Q_c={\rm diag.}\,\(\frac{2}{3},\,-\frac{1}{3}\) $ is the charge matrix and $\tau_i$ are the Pauli matrices. The quantity \eqref{Qg} can be obtained from the usual topological charge replacing the partial derivatives by the covariant derivatives \cite{Piette:1997ny,Livramento:2023keg}. The $U(1)$ gauge covariant generalization of the usual topological charge \eqref{chargemain0}
associated with Skyrme-Maxwell theory \eqref{Lag} and the gauge transformation \eqref{gauge} is defined by subtracting the magnetic contribution
$Q_{\rm mag}=\int d^3 x\frac{i\,g}{32\,\pi^2}\,\(\varepsilon_{ijk}\,{F}_{jk}\)\,\tr \(\left\{\tau_3,\,\pa_i U\right\} \, U^{-1}\)$ from \eqref{Qg}, i.e.
\be
Q\equiv Q_g-Q_{\rm mag} = \int d^3x\, q + \int d^3x\, \pa_i \Lambda_i\, ,
\label{Q}
\ee
where the surface term
depends on the boundary conditions imposed on the fields \cite{Piette:1997ny,Cork:2021ylu},
\be
\Lambda_i = -\frac{g}{4\,\pi^2}\,\varepsilon_{ijk}\,{A}_j\,\(\phi_3\,\pa_k\phi_0 - \phi_0\,\pa_k\phi_3\,\)
\label{lambdaapp}~.
\ee
In the Abelian Skyrme-Maxwell model \re{Lag} the flux of $\Lambda_i$ is vanishing,
$Q=Q_T$ \cite{Piette:1997ny,Livramento:2023keg}.

The stress-energy tensor which follows from \re{Lag0} is
\begin{eqnarray}
T^{\mu\nu}=T^{\mu\nu}_{(M)}+T^{\mu\nu}_{(S)}\label{T}~,
\end{eqnarray}
where the electromagnetic contribution  is
\begin{eqnarray}
T^{\mu\nu}_{(M)}=-2\,F^{\mu\sigma }\,F^{\nu}_{\:\:\sigma}+\frac{\eta^{\mu\nu}}{2}\,F_{\alpha\beta}\,F^{\alpha\beta},
\end{eqnarray}
and the  stress-energy tensor of the U(1)-gauged Skyrmions is
\begin{eqnarray}
\nonumber T^{\mu\nu}_{(S)}&=&2\,\left[ D^\mu \phi_a\,D^{\nu}\phi^a-\(D^{[\mu} \phi^a\,D^{\alpha]} \phi^b\)\,\(D^{[\nu} \phi_a D_{\alpha]}\phi_b\)\right]  \\
&&-\eta^{\mu\nu}\,\(\( D_\alpha \phi_a \)^2-\frac12 \( D_{[\alpha} \phi_a \,D_{\beta]} \phi_b \)^2- 2\,m^2\,(1-\phi_0)\) \,  \label{stress}.
\end{eqnarray}

For stationary solutions in the static gauge the Lagrangian and Hamiltonian of the Skyrme-Maxwell model can be
written, respectively, as
\be
{\cal L}_{{\rm static}}=-{\cal H}_{1} + {\cal H}_{2} \, ,
\qquad\qquad\qquad{\cal H}_{{\rm static}}={\cal H}_{1} + {\cal H}_{2} \, ,
\label{staticL}
\ee
where ${\cal H}_1$ and ${\cal H}_2$ are non-negative terms given by
\be
\begin{split}
%\label{Lagstatic}
{\cal H}_{1}=& \frac{1}{2}\,\mid F_{ij}\mid^2+\mid D_i \phi_a \mid^2
+\frac12 \mid D_{[i} \phi_a \,D_{j]} \phi_b \mid^2+ 2\,m^2\,(1-\phi_0)\, ,\\
{\cal H}_2=&\mid \pa_i A_0\mid^2+g^2\,A_0^2\,M_\phi^2,\quad {\rm with }
\quad M_\phi^2\equiv \left[\(1+\mid\pa_i \phi_A\mid^2\)\,\mid \phi_\alpha \mid^2+\frac{1}{4}\,
\left|\pa_i \(\mid \phi_A\mid^2\)\right|^2\right] \, ,\label{H12}
\end{split}
\ee
where $\mid \pa_iA_0 \mid^2=\pa_iA_0\,\pa_iA_0$, and so on. Clearly, the term ${\cal H}_2$ yields the Gauss law. Imposing the Coulomb gauge, $\pa_i A_i=0$,
we can write the static
Maxwell equations of the system
\be
\pa_j^2 A_0 = g^2\,M_\phi^2 \,A_0 ,\quad \pa_j^2 A_i = -g\,\varepsilon_{\alpha\beta}\,\phi_\beta\,\left[\(1+ D_j \phi_a D_j \phi_a \)\,D_i \phi_\alpha  - \(D_j \phi_a\,D_j \phi_\alpha\)\,D_i \phi_a
\right]\,.\label{A0eq}
\ee

\section{$U(1)$ gauged Skyrmions}
\label{sec:3}

Hereafter we consider the case of vanishing electrostatic potential $A_0(\vec{r})=0$,
so, the Gauss law  is satisfied trivially. Consequently, the electric charge vanishes and
${\cal H}_2=0$. The total static energy of the gauged Skyrmions can be written as
\be
E = E_2+E_4 +E_0 + E_{{\rm em}} \label{energy_tot}\, ,
\ee
where
\begin{eqnarray}
E_{2}&=&\frac{1}{12\, \pi^2}\,\int d^3x\,\mid D_i \phi_a \mid^2,\qquad\qquad\qquad \,\, E_4=\frac{1}{12\, \pi^2}\,\int d^3x\,\frac12 \mid D_{[i} \phi_a \,D_{j]} \phi_a \mid^2, \nonumber\\ E_0&=&\frac{1}{12\, \pi^2}\,\int d^3x\, 2\,m^2\,(1-\phi_0), \qquad\qquad  E_{{\rm em}}=\frac{1}{12\, \pi^2}\,\int d^3x\,\frac12 \, \mid F_{ij} \mid^2  \label{es}
\,,
\end{eqnarray}
and we make use of the standard normalization of the energy functional by the factor $12\,\pi^2$.

The critical points of the total energy functional \re{energy_tot}
should satisfy the arguments of the Derrick's theorem \cite{Derrick:1964ww}.
Under a  scale transformation $x_i \rightarrow x'_i= \lambda\,x_i$,
the scalar fields $\phi_a$ and the four-potential transform as
$\phi_a \rightarrow \phi_a $ and $A_\mu \rightarrow A_\mu'\(\vec{x}'\)=\lambda^{-1}A_\mu\(\vec{x}\)$. Thus,
\be
E\(\lambda\) = \lambda^{-1} \, E_2+\lambda \,\(E_4 + E_{{\rm em}}\)+\lambda^{-3} \,E_0 \, ,
\ee
and
\be
\pa_\lambda^2 E(\lambda=1)=2\,\left[E_2+6\,\,E_0\right] \geq 0\, . \nonumber
\ee
The corresponding  virial identity follows from the condition $\pa_\lambda E(\lambda=1)=0$,
it gives $E_2+3\,E_0=E_4+E_{{\rm em}}$, which
can also can be obtained from \eqref{stress} using the von Laue stability condition
$\int d^3x\, T^i_i=0$ \cite{Laue:1911lrk,Bialynicki-Birula:1993shm}.

There is a variety of classical multisoliton solutions of the usual Skyrme model
constructed numerically over last three decades, see \cite{Manton:2004tk,Manton:2022}.
As the electromagnetic interaction is decoupled, the  basic unit charge Skyrmion represent
a spherically symmetric hedgehog \cite{skyrme,Skyrme:1962vh}.  Skyrmions of higher topological degrees may
possess much more complicated symmetries, the shape of the multisoliton configuration
strongly depends on the particular choice of the potential and corresponding character of
asymptotic decay of the fields.
Coupling to the electromagnetic field yields an additional interaction,
our full 3d numerical simulations confirmed that
$U(1)$ gauged Skyrmions of topological degrees $Q=1,2$ are axially symmetric \cite{Piette:1997ny,Livramento:2023keg}.
These Skyrmions can be constructed using a general reduced Ansatz
\cite{Krusch:2004uf,Battye:2005nx,Radu:2005jp,Ioannidou:2006nn}:
\begin{eqnarray}
\label{ans_ax}
 \phi_1+i\, \phi_2=\psi_1(r,\theta)\, e^{i\, n\,\varphi}\ , \qquad
 \phi_3=\psi_2(r,\theta) \ , \qquad \phi_0=\psi_3(r,\theta) \ ,
\end{eqnarray}
where $n$ is an integer which counts the winding of the unit vector field $\vec \psi=(\psi_1,\psi_2,\psi_3)$, $|\psi|^2=1$
in equatorial plane. In a more general case,  the topological charge of the axially-symmetric Skyrmions is given by the product of two integers,
$Q=mn$, there $n$ is the phase winding number \eqref{ans_ax} and $m$ is a number of twists along the toroidal fundamental cycle of the torus, see e.g. \cite{Gudnason:2014jga,Shnir:2009ct}.
For all solutions we are discussing here, the number of twists $m=1$, so $Q=n$ \cite{Krusch:2004uf,Battye:2005nx,Radu:2005jp}.
Note that the Ansatz \re{ans_ax} also can be used to construct saddle point
solutions of the Skyrme model, which represent chains of interpolating Skyrmion--anti-Skyrmions of topological degree $\pm n$
\cite{Krusch:2004uf,Shnir:2009ct}.

Since in the present study we do not consider electrically charged Skyrmions, the gauge field
is parameterized by the magnetic potential solely:
\begin{eqnarray}
\label{A}
A \equiv A_\mu dx^\mu=A_\varphi (r,\theta) d\varphi.
\end{eqnarray}
All four functions which parameterize the axially-symmetric Ansatz \eqref{ans_ax}, \eqref{A}
depend on the radial variable $r$ and the polar angle $\theta$.
Within this Ansatz, the Skyrme-Maxwell static Lagrangian \re{staticL}, normalized by the factor
$12\,\pi^2$, and the energy \re{energy_tot} become:
\begin{eqnarray}
L&=&  -\frac{1}{6\,\pi} \,\int d\theta \,dr\,r^2 \sin \theta\left[
{\cal F}
+{\cal L}_{2}
+{\cal L}_{4}
+2\,m^2\,(1-\psi_3)
\right] \, , \qquad\quad E=-L\, ,\label{lageff}\end{eqnarray}%
where
\begin{eqnarray}
&&
{\cal F} \equiv\frac{1}{r^2 \sin^2 \theta}
\left(
A_{\varphi,r}^2+\frac{A_{\varphi,\theta}^2}{r^2}\)
\, ,\qquad\qquad
{\cal L}_{2}\equiv
  \psi_{a,r}^2+\frac{\psi_{a,\theta}^2}{r^2}
+\psi_1^2
\,
\frac{(n+g\, A_\varphi)^2}{r^2\sin^2 \theta}
,
\\
&&
{\cal L}_{4}\equiv\frac{1}{r^2}
\left[
(\psi_{3,\theta} \psi_{2,r} -\psi_{2,\theta} \psi_{3,r} )^2
+(\psi_{2,\theta} \psi_{1,r} -\psi_{1,\theta} \psi_{2,r} )^2
+(\psi_{3,\theta} \psi_{1,r} -\psi_{1,\theta} \psi_{3,r} )^2
\right]\\
\nonumber
&&
{~~~~~~~~~}
+ \psi_1^2
\,
\frac{(n+g\, A_\varphi)^2}{r^2 \sin^2 \theta}
\,
\left(
\psi_{a,r}^2+\frac{\psi_{a,\theta}^2}{r^2}\right) \, .
\end{eqnarray}
Here a comma denotes partial differentiation, i.e. $A_{\varphi,r}\equiv \frac{\partial A_\varphi}{\partial r}$, $etc$.
The corresponding
static field equations can be obtained from the variation of the effective Lagrangian
\re{lageff} with respect to the functions parametrizing the Ansatz \re{ans_ax}, \re{A}.

It was pointed out \cite{Krusch:2004uf} that the solutions of the Skyrme model constructed via the reduced Ansatz \re{ans_ax}
have certain similarity with Hopfions in the Faddeev-Skyrme model
\cite{Faddeev-Hopf,Faddeev:1996zj,Gladikowski:1996mb}. Furthermore, in the $U(1)$ gauged Skyrme model,
time-dependent gauge transformations \eqref{gauge} are associated with isorotations of the scalar field
 \cite{Radu:2005jp}.
% \be \phi_1+i\,\phi_2\rightarrow e^{-i\,g\,\alpha}\,\(\phi_1+i\,\phi_2\),\qquad A_\mu \rightarrow A_\mu + \pa_\mu \alpha\label{gauge}~,\ee REPEATED, see eq (5)
Hence, one of the components of the pion triplet can be set to zero, effectively truncating the model to the
$U(1)$ gauged Faddeev-Skyrme-Maxwell theory \cite{Shnir:2014mfa,Samoilenka:2018oil}.
However, there still is topological difference between such configurations and axially symmetric Hopfions classified by $\pi_3(S^2)$. 
Below we will see that numerical simulations 
clearly confirm the distinct distribution of the fields of the axially symmetric Hopfions and the $U(1)$ gauged Skyrmions.

%%%%%%%%%%%%%%%%%%%%%%%%%%%%%%%%%%
\section{Numerical scheme }
\label{sec:4}
%%%%%%%%%%%%%%%%%%%%%%%%%%%%%%%%%%

To find stationary points of the energy functional \re{energy_tot} we
use a suitable combination of the gradient descendent method, implemented for solving the system
of Euler-Lagrange equations for the scalar field,
and the  Newton-Raphson method for solving  the Maxwell equation for the magnetic potential.
Generally, we do not impose any restrictions of symmetry, however,
in order to check the consistency of our numerical calculations, in case of the axially-symmetric solitons
with topological charges $Q=1,2$,
we compare the results of full 3d simulations with reduced solutions of the truncated system \re{lageff}.

Most of our simulations are performed on  cubic grids of $(121)^3$ points with a lattice
spacing $\Delta x =0.1$ and adaptive step size. We impose Dirichlet boundary conditions $\phi^a=(1,0,0,0)$
on the boundary of the cube and make use of a 4th order finite difference scheme. The boundary conditions on the magnetic potential are
$A_\phi(0)=A_\phi(\infty)=0$.
For avoid local minima of the energy functional \re{energy_tot} for all charges from 1 to 5,
we accept only results which provide the same global minimiser of $E$
for different initial guesses generated by the
rational map approximation, the product Ansatz and the axially symmetric Ansatz \eqref{ans_ax}, \eqref{A}.\footnote{The rational map is a powerful Ansatz to constructing very symmetric initial field configuration which leads to the global energy minimizer for the Skyrme model. However, in our simulations, for $Q=4$ and $g>0$ these initial settings did not lead to the global minimizer, which can be obtained by starting with four $Q=1$ Skyrmions or two $Q=2$ Skyrmions combined through the product Ansatz.}
We also check the results of our fully three-dimensional numerical simulations with those
obtained by solving the reduced two-dimensional system of equations for
axially-symmetric solutions of degrees one and two. In the latter case
we put the system on a square domain with  $70^2$ points
and map the infinite interval of the variable $r$ onto the compact radial coordinate $x=\frac{r}{2+r} \in [0:1]$.

The gradient flows of the scalar field $EL_{\phi_a}$ and the gauge field $EL_{A_a}$  are
\begin{eqnarray}
EL_{\phi_a} &\equiv & \frac{1}{2}\,\(\frac{\delta L_{{\rm static}}}{\delta \phi_a}- \phi_a\,
\sum_{b=0}^3\,\phi_b\frac{\delta L_{{\rm static}}}{\delta \phi_b}\)\, ,
\label{equations}\\
EL_{A_i} &\equiv & \pa_j^2 A_i + g\,\varepsilon_{\alpha\beta}\,\phi_\beta\,\left[\(1+ D_j \phi_a D_j \phi_a \)\,D_i \phi_\alpha  - \(D_j \phi_a\,D_j \phi_\alpha\)\,D_i \phi_a
\right]\,, \quad EL_{A_0} \equiv 0\,,\quad\label{equations2}
\end{eqnarray}
respectively. Energy minima of the system correspond to $EL_{\phi_a}=0,~EL_{A_i}=0 $,  the accuracy check of our simulations is to
evaluate the root-mean-square deviation (RMSD) of the Euler-Lagrange gradients
\be
\Delta_E = \sqrt{\frac{1}{N}\,\sum_{a=0}^4 \(\mid EL_{\phi_a} \mid^2+ \mid EL_{A_a} \mid^2\)} \, ,
\label{rmsd}
\ee
where $N$ is the number of points of the lattice. Typically, for our numerically-determined minima on a cubic lattice,
we obtain $\Delta_E\sim 10^{-8}$.\footnote{The accuracy for non-toroidal $Q=4$ gauged Skyrmions ($g=0.1,...,\,0.6$) decreases to $\Delta_E <6 \times 10^{-5}$. In addition, in such a case the maximum values of the modulus of the gradients \eqref{equations} over all the lattice points satisfies ${\max}\,\mid EL_a\mid<8 \times 10^{-4}$ and ${\max}\,\mid EL_{A_a}\mid < 10^{-7}$.} As a further check of our numerics, we calculate numerically the topological charge
$Q_n$ through \re{chargemain0} and compare it to its true integer value $Q$,
$\mid 1-Q_n/Q\mid \sim 3\times 10^{-4}$. We also checked that the virial relation is satisfied within $1-2\%$
accuracy.
In the case of the 2d simulations, the  errors are reduced further by an order of magnitude.

%%%%%%%%%%%%%%%%%%%%%%%%%%%%%%%%%%
\section{Numerical results}
\label{sec:5}
%%%%%%%%%%%%%%%%%%%%%%%%%%%%%%%%%%
\begin{figure}[h]
\begin{center}
\includegraphics[scale=0.11]{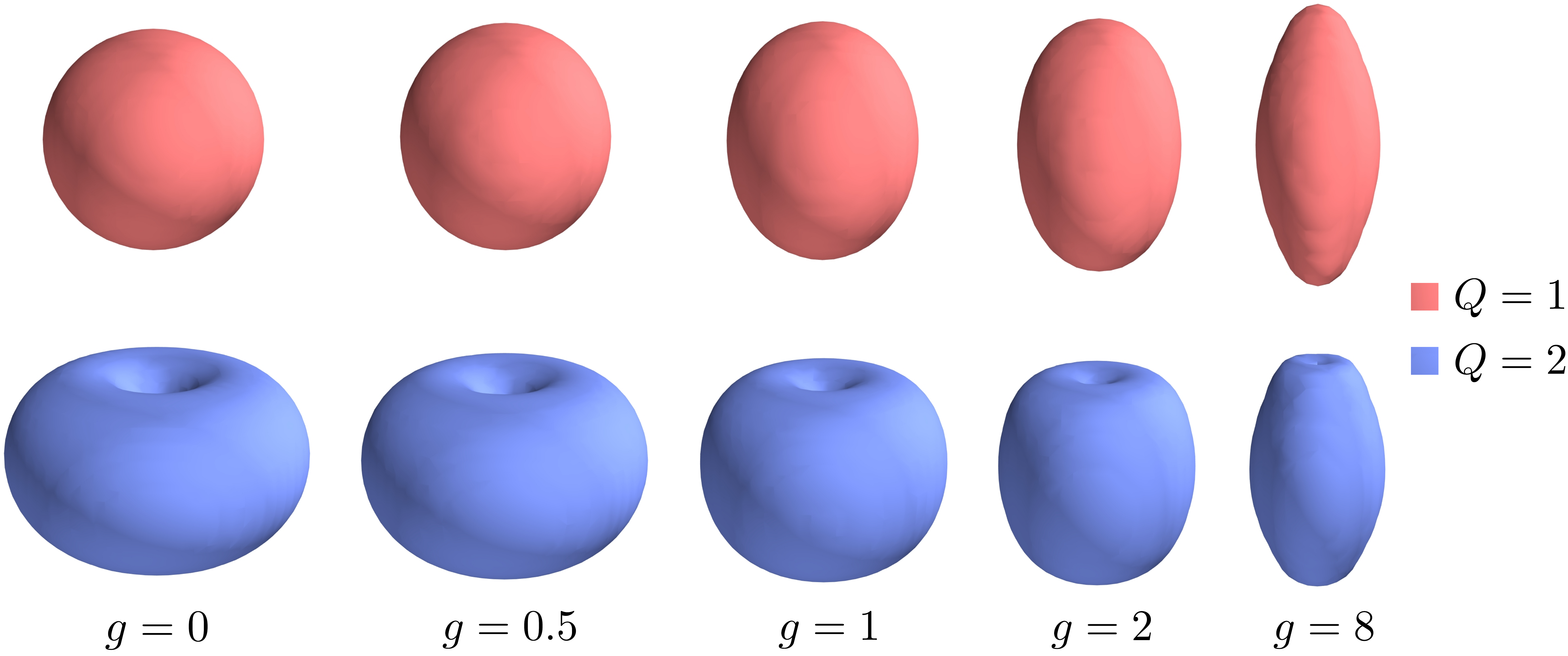}
\end{center}
\caption{Isosurfaces of the total energy density
distributions of the gauged Skyrmions with $Q=1$ (top row) and $Q=2$
(bottom row) are displayed for some values of $g$ for
${\cal E}=0.08$. In each row all of the solutions are plotted at the same scale.
}
\label{fig1}
\end{figure}

\begin{figure}[h]
\begin{center}
\includegraphics[scale=0.25]{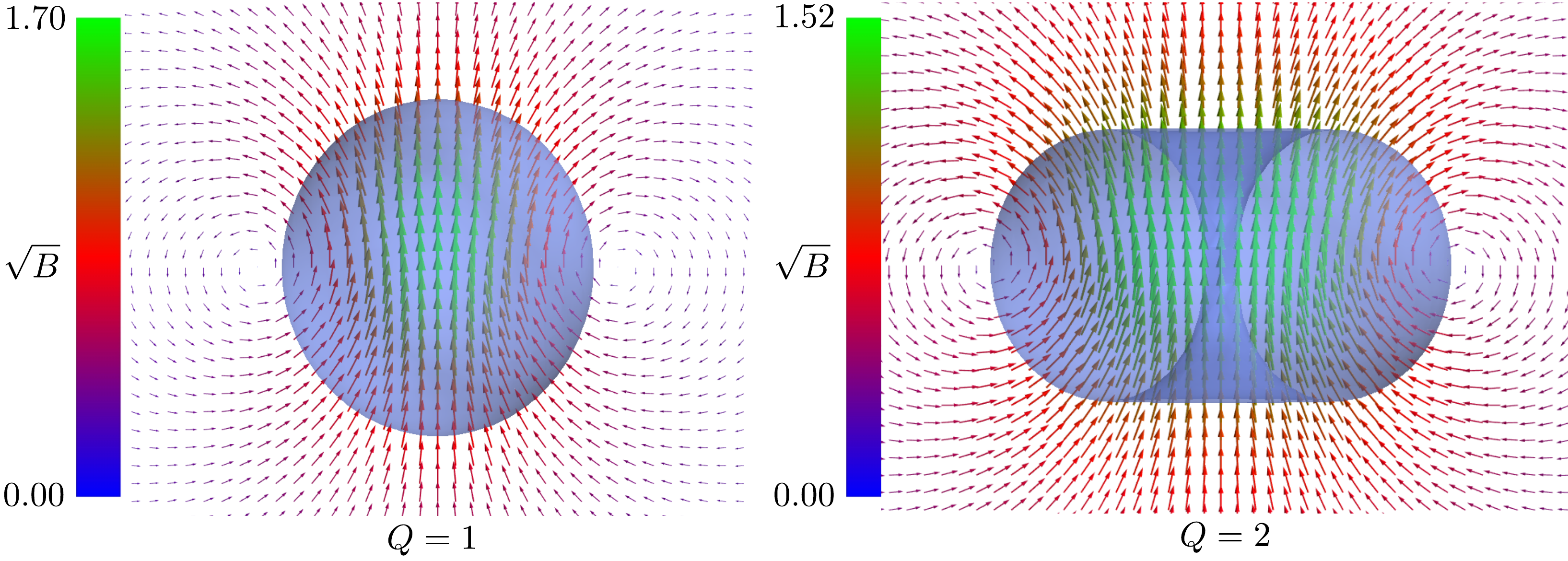}
\end{center}
\caption{Distribution of the magnetic field $\vec{B}$ in the $x-z$ plane is dispayed for
the axially-symmetric gauged Skyrmions of topological degrees one (left plot) and two (right plot) for
$g=0.5$. The length of the plotted vectors is proportional to the magnitude of $\vec{B}/\sqrt{B}$, where $B\equiv \mid \vec{B}\mid$. The translucent blue surface is a level set of
constant total energy density for ${\cal E}=0.1$.
}
\label{fig2}
\end{figure}

\begin{figure}[h]
\begin{center}
\includegraphics[scale=0.325]{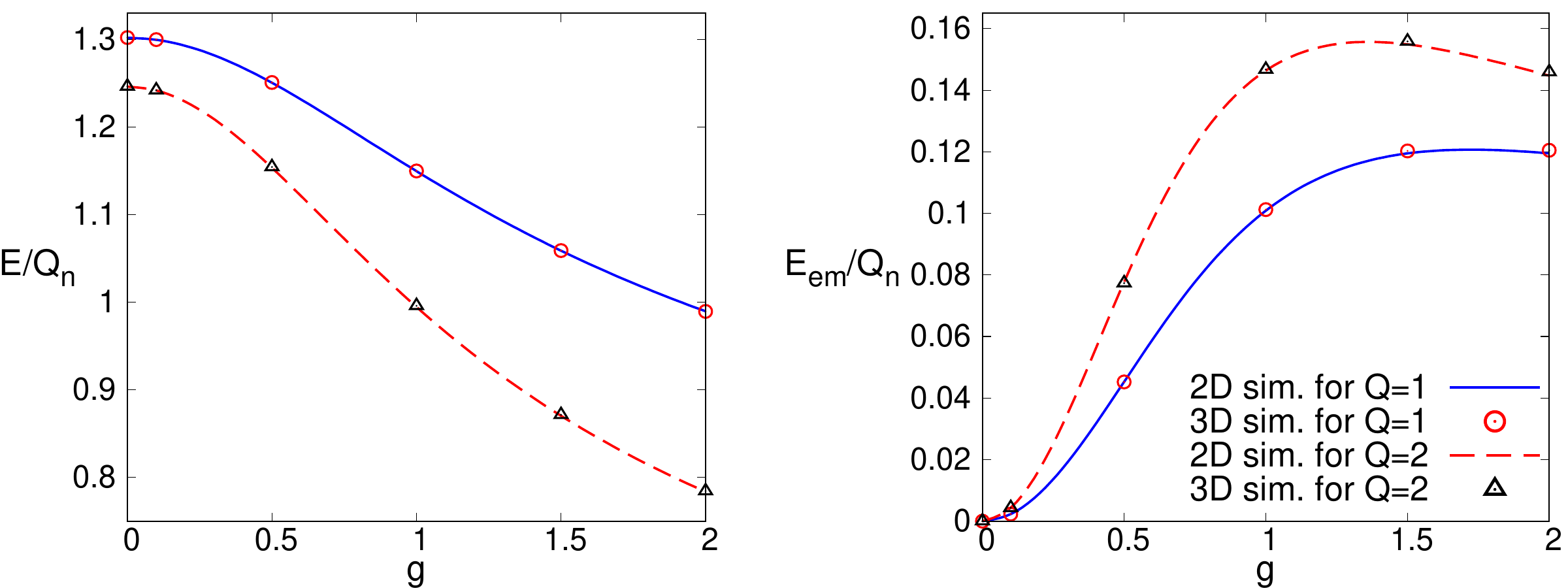}
\end{center}
\caption{The total energy $E$ and the electromagnetic energy $E_{{\rm em}}$ of the axially-symmetric Skyrmions
with $Q=1,\,2$, per unit charge.
The curves and the dots corresponds to the results obtained using
reduced axially symmetric Ansatz \re{ans_ax}, \re{A} and with fully three-dimensional numerical simulations,
respectively.
}
\label{fig3}
\end{figure}

First, we considered axially-symmetric gauged Skyrmions with topological charges $Q=1,\, 2$.
As the gauge coupling gradually increases from $g=0$, the energy of the configuration decreases\footnote{However, the binding energy of the gauged Skyrmions per unit charge increases \cite{Piette:1997ny}.
} since the
local magnetic flux is formed in the equatorial plane and core of the soliton shrinks, see Figs.~\ref{fig1}, \ref{fig2}.
The position of the circular flux is associated with a minimum of the magnetic potential $A_\varphi$.
The electromagnetic energy  is initially increasing,
however its contribution starts to decrease as $g$ becomes larger than $g\sim 1$, see Fig.~\ref{fig3}.
The total magnetic flux of the
configuration is zero, it is coupled to the Skyrmion providing its magnetic moment.

As the coupling becomes stronger, the magnetic potential develops a plateau in the equatorial plane,
where $gA_\varphi+n \sim 0$. Here the integer $n$ is to the winding number
of the unit vector field $\vec \psi$ \re{ans_ax}. This corresponds to
a string of magnetic flux through the center of the configuration. The flux is non-topologically
quantized in units of $2 \pi$ and carries $n$ quanta.
It is known, such a situation is common for all $U(1)$ gauged solions, like
Hopfions in the Faddeev-Skyrme-Maxwell model \cite{Shnir:2014mfa}, gauged planar
Skyrmions \cite{Gladikowski:1995sc,Shnir:2015twa,Samoilenka:2015bsf}
and gauged $O(3)$ lumps \cite{Schroers:1995he}.

\begin{figure}[H]
\begin{center}
\includegraphics[scale=0.105]{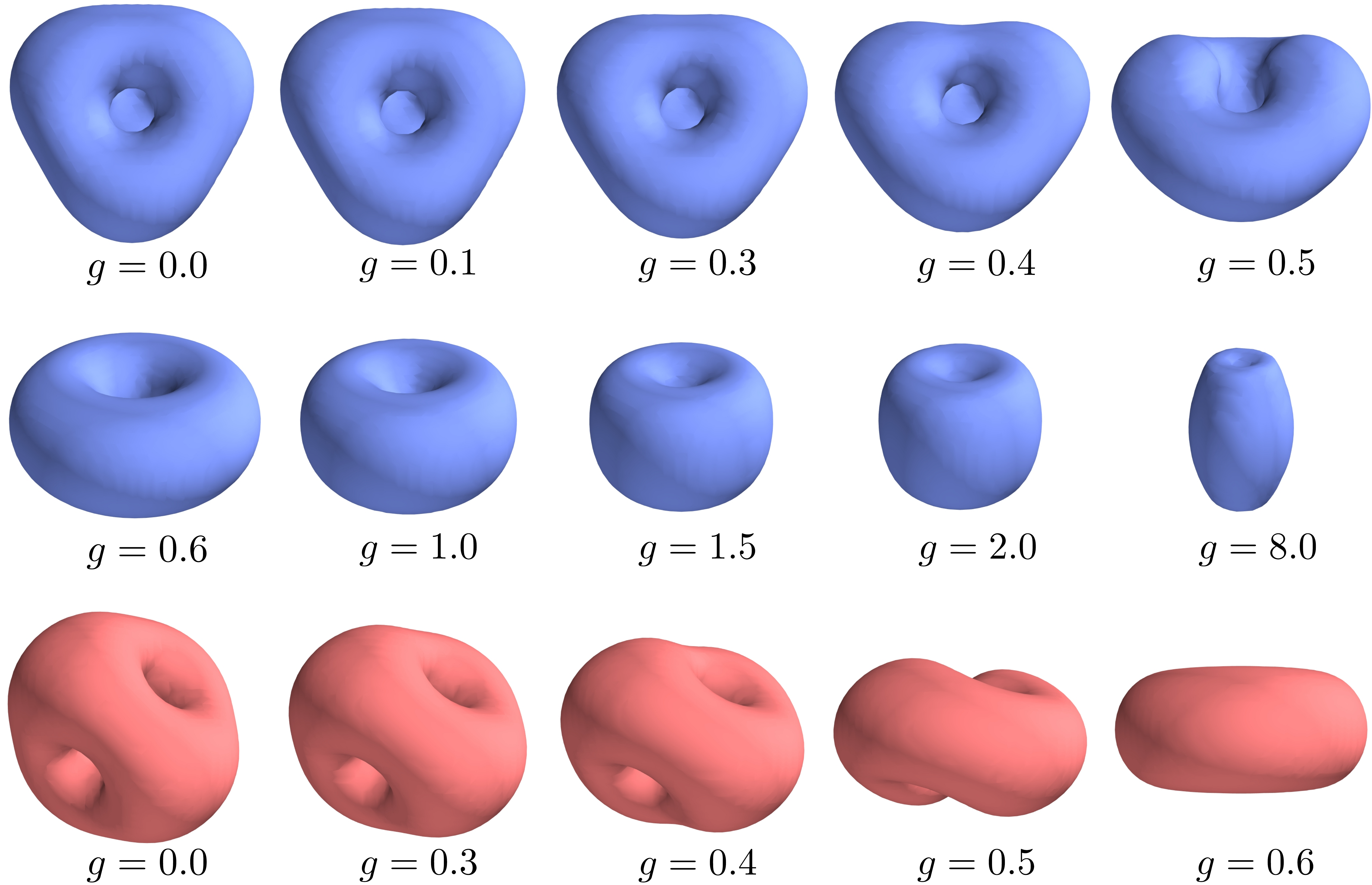}
\end{center}
\caption{Isosurfaces of the total energy density
of the gauged Skyrmion of degree $Q=3$ for some set of values of $g$ at
${\cal E}=0.08$. The bottom row displays the side view of these Skyrmions.
In each row all of the solutions are plotted at the same scale.}
\label{fig4}
\end{figure}
\begin{figure}[b]
\begin{center}
\includegraphics[scale=0.33]{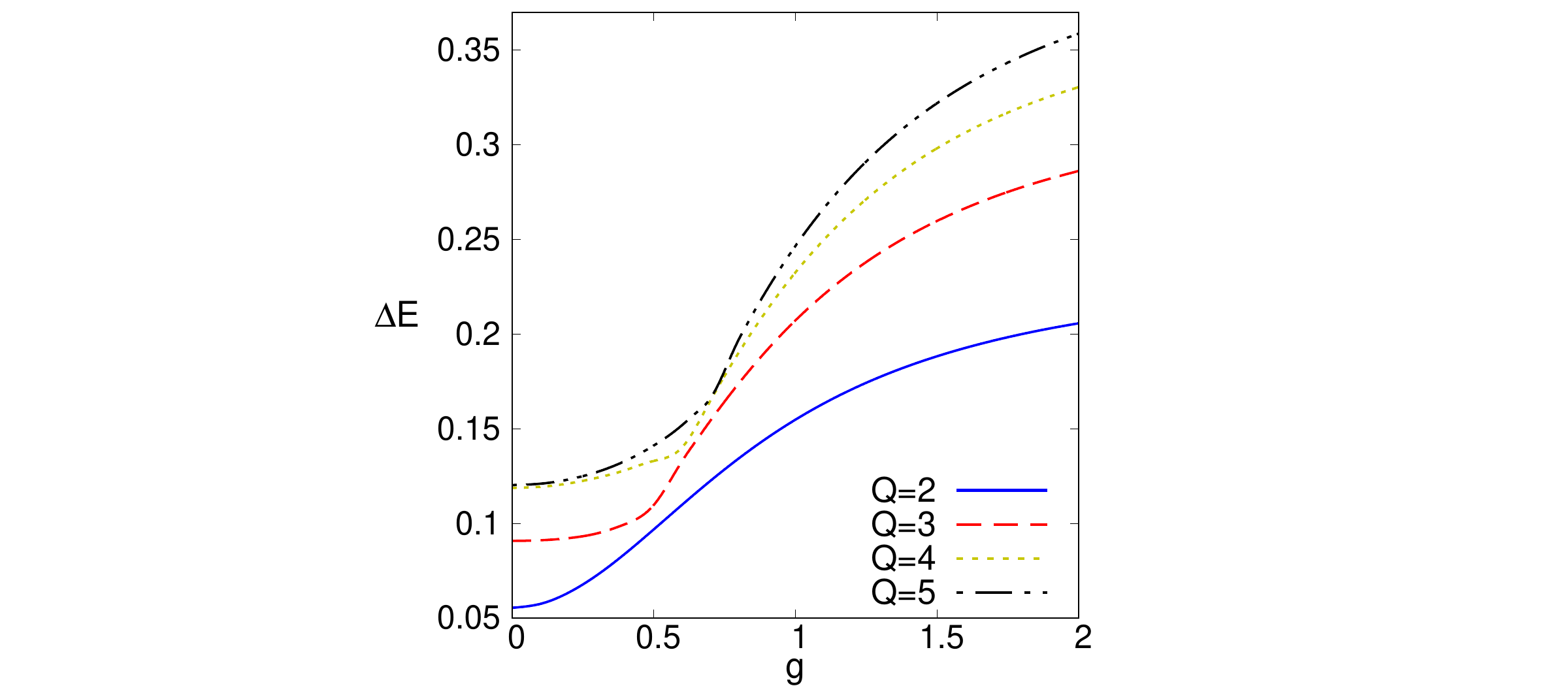}
\end{center}
\caption{
The binding energy per topological charge unit $\Delta E(g) \equiv E_{Q=1}(g) - E_{Q}(g)/Q$ of the Skyrmions
with $Q=2-5$. The values of the energies are shown in the lines plotted in Figs. \ref{fig3} and \ref{fig8}.
}
\label{binding}
\end{figure}

The distribution of the magnetic field of the $Q=3$ configuration is different from the case of axially-symmetric Skyrmions,
see Fig.~\ref{fig5}. In the decoupled limit $g=0$ the total energy density distribution is tetrahedrally
symmetric, appearance of the magnetic flux which encircles the Skyrmion breaks this symmetry. The total magnetic
flux is zero, the four lumps at the vertices of the tetrahedron are bounded to local magnetic fluxes, as displayed in
Fig.~\ref{fig5}. As  the gauge coupling increases, these local fluxes move towards the center of the configuration, they merge
into a single return flux penetrating the center of the gauged $Q=3$ Skyrmion as $g\ge 0.6$.

Turning now to solitons of higher degrees, we consider gauged Skyrmions with $Q=3-5$.
In the ungauged limit the $Q=3$
Skyrmion has tetrahedral symmetry, coupling to the magnetic flux breaks it down.
Fig.~\ref{fig4} displays the corresponding isosurfaces of
the total energy densities for the gauged Skyrmions  for
some set of values of the gauge coupling $g$ up to $g=8$.
Clearly, geometrical shape of the $Q=3$ soliton varies from a regular tetrahedron at $g=0$ to
a torus at $g\le 0.6$.  Note that at $g\approx 0.5$ the energy density distribution of a gauged Skyrmion
takes the shape of a bent torus, which is exactly the shape of the ground state Hopfion with topological charge 3.
As said above, there is an evident similarity between the axially-symmetric solution of the Skyrme-Maxwell theory and
Hopfions, the curve of positions of minima of the component $\phi_0$ correspond to the position curve of the Hopfion.
Interestingly, this similarity also persists for the charge 3 gauged solitons which do not possess axial symmetry.

Considering the binding energy of the gauged Skyrmions we confirm confirm earlier results that the  
binding energy of the gauged Skyrmions per unit charge increases \cite{Piette:1997ny}. Fig.~\ref{binding} displays the binding energy 
of the $Q=2-5$ gauged Skyrmions vs gauge coupling $g$. Evidently, it moves further from the topological bound as $Q$ increases.

Physically, the transformation of a Skyrmion from a configuration with discrete symmetry to the toroidal state, corresponds to
a  phase transition induced by the axially-symmetric magnetic field. This effect is similar to  transitions to toroidal Skyrmions in the Skyrme model with symmetry breaking potential   \cite{Gudnason:2014jga}.
\begin{figure}[H]
\begin{center}
\includegraphics[scale=0.11]{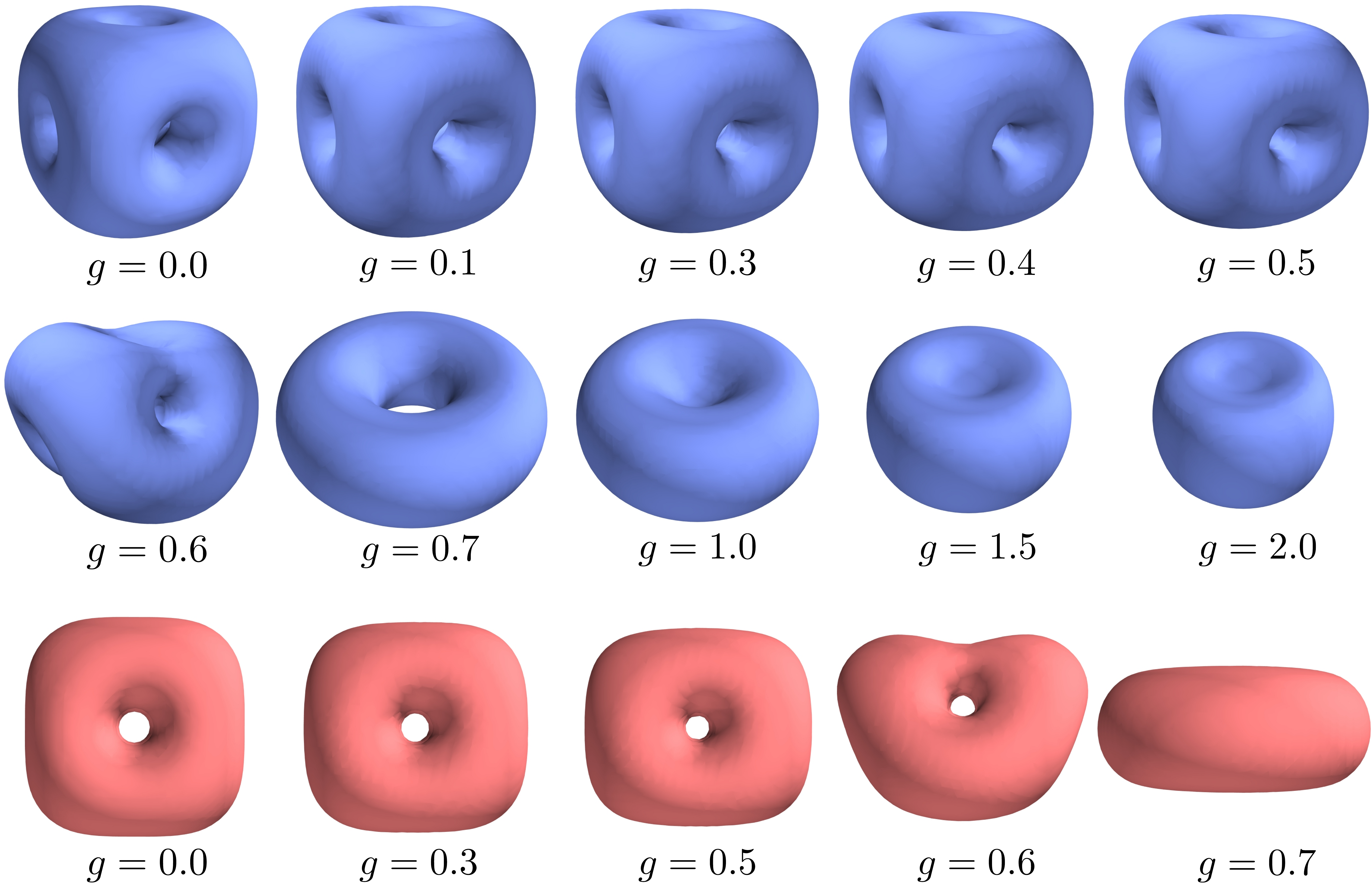}
\end{center}
\caption{Isosurfaces of the total energy density
of the gauged Skyrmion of degree $Q=4$ for some set of values of $g$ at
${\cal E}=0.08$. The bottom row displays the side view of these Skyrmions.
In each row all of the solutions are plotted at the same scale.
}
\label{fig6}
\end{figure}
\begin{figure}[H]
\begin{center}
\includegraphics[scale=0.11]{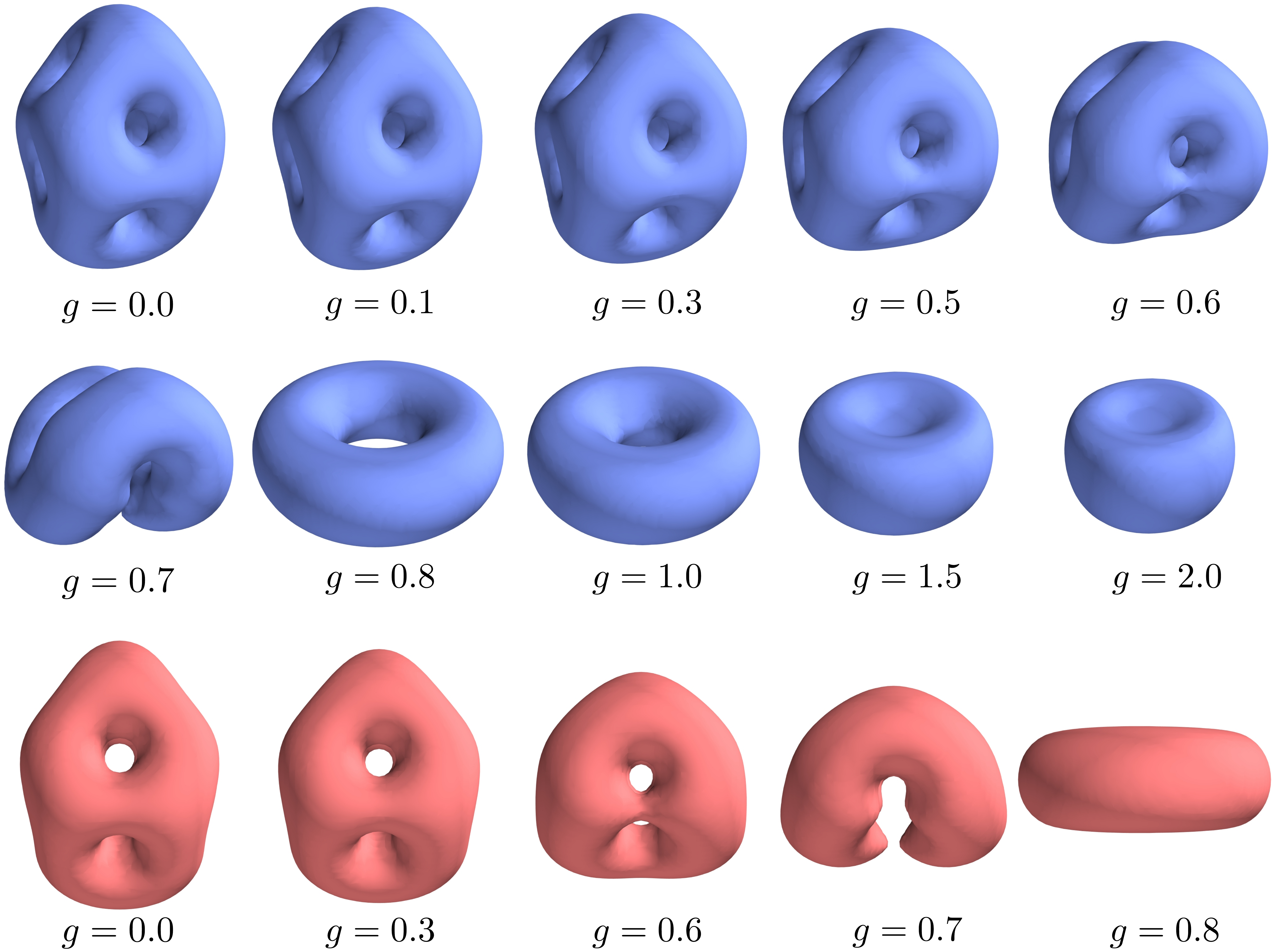}
\end{center}
\caption{Isosurfaces of the total energy density
of the gauged Skyrmion of degree $Q=5$ for some set of values of $g$ at
${\cal E}=0.08$. The bottom row displays the side view of these Skyrmions.
In each row all of the solutions are plotted at the same scale.
}
\label{fig7}
\end{figure}

\begin{figure}[b]
\begin{center}
\includegraphics[scale=0.228]{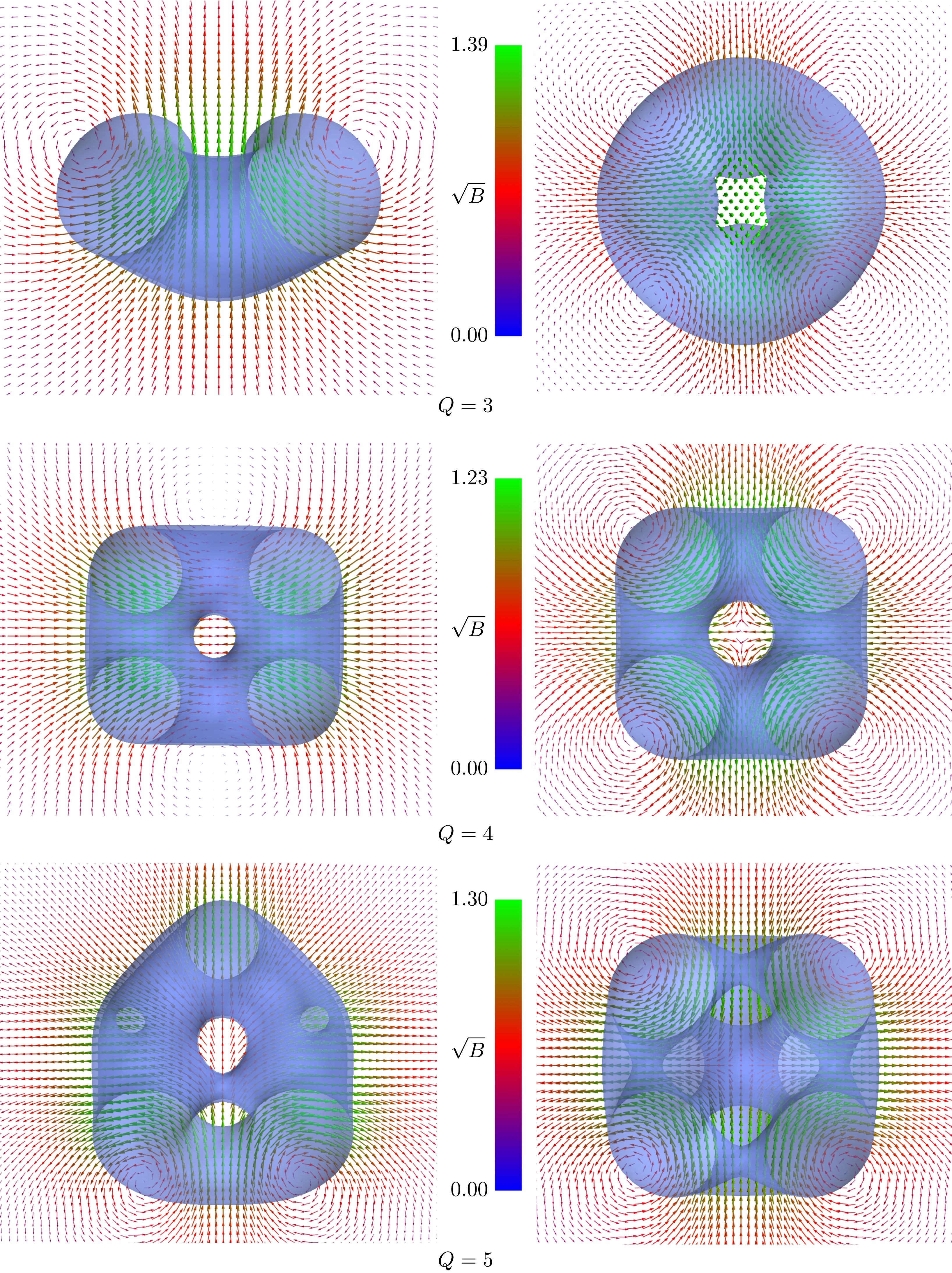}
\end{center}
\caption{
Distributions of the magnetic field $\vec{B}$ in the $x-z$ plane (left plots) and in the $x-y$ plane (right plots) for the gauged Skyrmions of topological degrees $3-5$ for
$g=0.5$.  The length of the plotted vectors is proportional to the magnitude of $\vec{B}/\sqrt{B}$, where $B\equiv \mid \vec{B}\mid$. The translucent blue surface is a level set of
constant total energy density for ${\cal E}=0.1$.
}
\label{fig5}
\end{figure}

\begin{figure}[b]
\begin{center}
\includegraphics[scale=0.34]{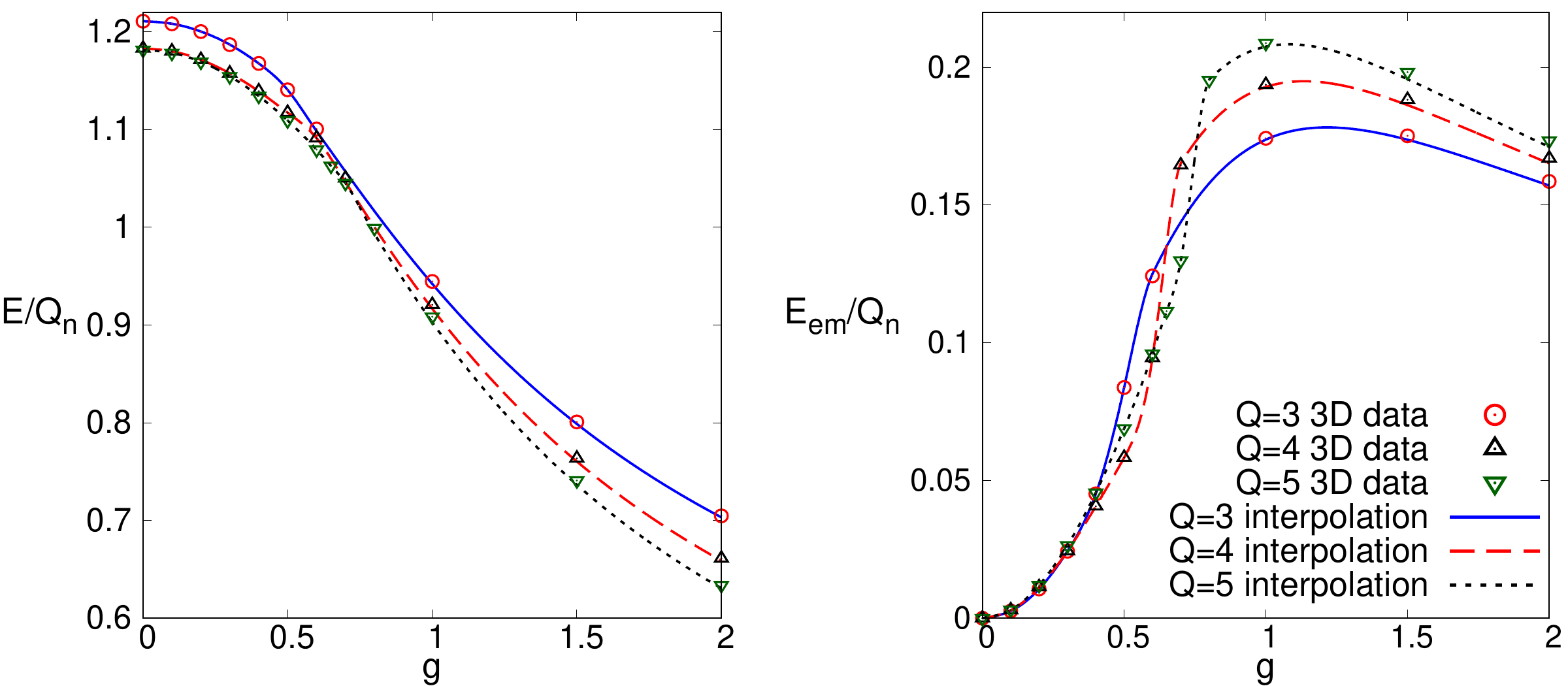}
\end{center}
\caption{
The total energy $E$ and the electromagnetic energy $E_{{\rm em}}$ of the  Skyrmions
with $Q=3-5$, per unit charge. The dots correspond to the values obtained through fully three-dimensional
numerical simulations, the curves are plotted using cubic interpolation and the data of the
fully three-dimensional simulations for $g \leq 0.5, \, 0.6,\,0.7$ for respectively $Q=3,\, 4, \,5$, and the results obtained inside the axially symmetric Ansatz for $g \geq 0.6, \, 0.7,\,0.8$, respectively.
}
\label{fig8}
\end{figure}

Numerical simulations show that qualitatively the same pattern appears for the gauged Skyrmions of higher degrees, see
Figs.~\ref{fig6}, \ref{fig7}, \ref{fig8}. The usual $Q=4$ Skyrmion for $g=0$ is octahedrally symmetric, the ungauged
$Q=5$ configuration can be approximately described by a bounded system of $4+1$ Skyrmions.

As the gauge coupling increase, the contribution of the electromagnetic energy  is  increasing at first,
it is maximal as $g\sim 1$, see Fig.~\ref{fig8}. This effect is related with deformations of the Skyrmions and formation of
magnetic flux.

The structure of the magnetic field follows the pattern above,
in the weak coupling regime there is a circular magnetic flux tubes, which encircles
the Skyrmions of degrees $3-5$. This flux is linked to the four return local fluxes, as seen in the right
column of Fig.~\ref{fig5}, where the circular magnetic field crosses the isosurface of total energy density of ${\cal E}=0.1$. Further increase of the gauge coupling drives the Skyrmions towards an axially-symmetric
solution bounded with a toroidal magnetic field with a flux along the symmetry z-axis
of the configuration, and the circular magnetic flux orthogonal to the $x-z$ plane, analogously to what can be seen for $Q=2$ on the right plot in Fig. \ref{fig2}. The
transformation to the toroidal $Q=4$ and $Q=5$ Skyrmions takes place around $g=0.7$ and $g=0.8$, respectively. Clearly, the behavior of the magnetic field is more complex for non-axially symmetrical gauge Skyrmions, its magnitude tends to be stronger inside the core of the Skyrmion. The Fig.~\ref{fig5} displays the magnitude and direction of the magnetic field for $g=0.5$.

\begin{figure}[t]
\begin{center}
\includegraphics[scale=0.29]{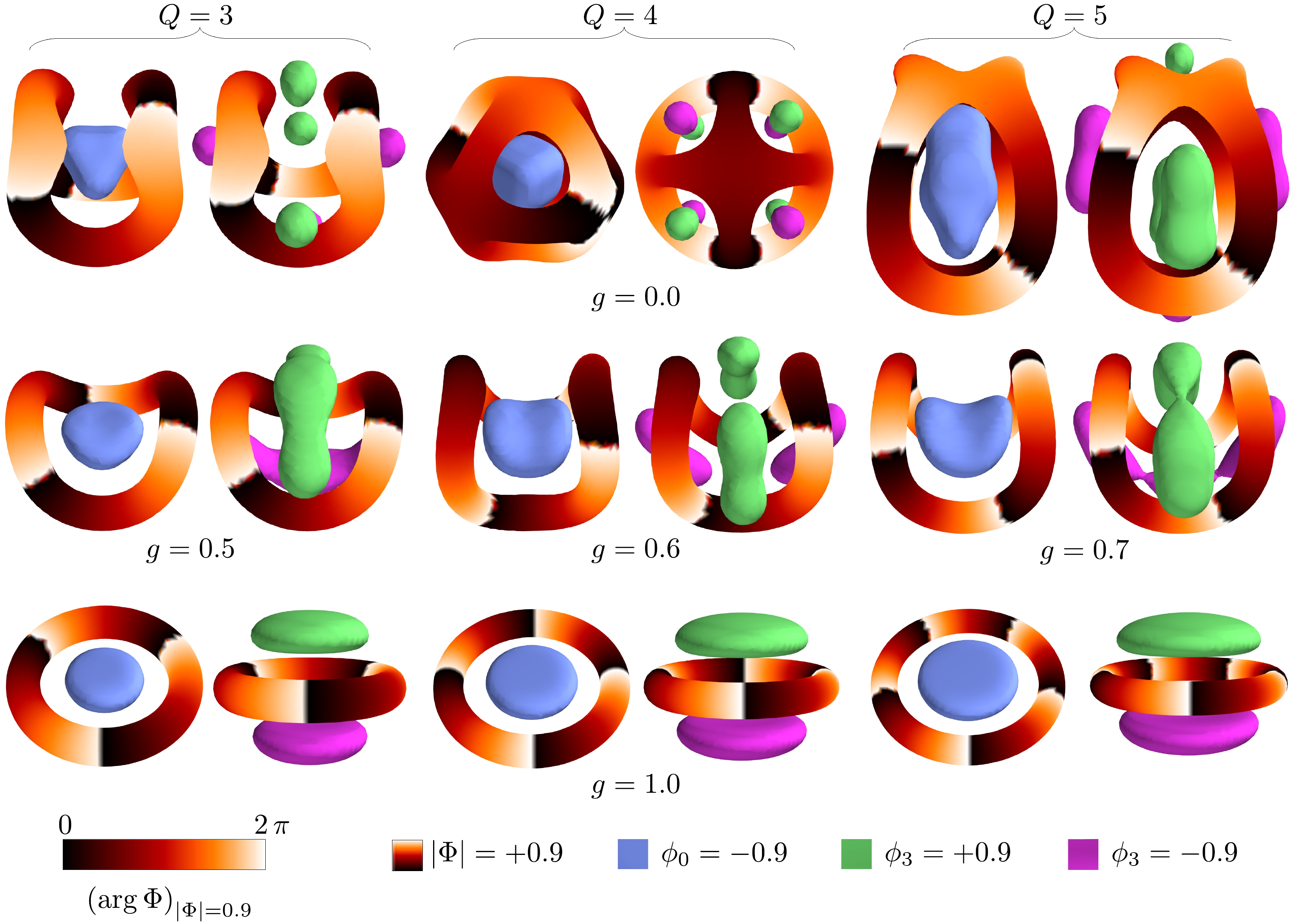}
\end{center}
\caption{
Isosurfaces of the field components $\phi_0=- 0.9,\,\phi_3=\pm 0.9$ and $\mid \Phi\mid =0.9$, where $\Phi\equiv \phi_1+ i\,\phi_2$, of the $Q=3,\,4,\,5$ gauged Skyrmions for some values of $g$.
The orientation in isospace is
visualized using the field colouring scheme
for the argument
${\rm arg} \(\Phi\) \in [0,\,2\pi)$.
}
\label{fig:Q=3-5}
\end{figure}

In Fig.~\ref{fig:Q=3-5} we illustrate transition to toroidal gauged Skyrmions. Here, we visualize the field configurations by plotting isosurfaces of the field components $\phi_0=-0.9$, $\phi_3=\pm 0.9$ and $\mid\Phi\mid = 0.9$, where $\Phi\equiv \phi_1+ i\,\phi_2$, for some set of values of the gauge coupling $g$. The colouring scheme displays
phase of the component $\Phi$. This also allows us to compare the
toroidal Skyrmions with Hopfions, the tube-like isosurface $\phi_0=-0.9$ can be set into correspondence with the position curve of the soliton, it is defined as the set of points where
the field $\phi_0$ is as far as possible from the boundary vacuum value $\phi_0=1$.

Recall that the topological charge of the toroidal Skyrmions \re{chargemain0} is given by the product of two integers,
$Q=mn$, were $n$ is the phase winding number and $m$ is a number of twists along the fundamental cycle of the torus.

However, the  solutions of the gauged Skyrme model in the strong coupling limit are different from the conventional axially-symmetric Hopfion classified by the Hopf invariant \cite{Gladikowski:1996mb,Kobayashi:2013xoa}, the field components of the toroidal gauged Skyrmions dispayed in Fig.~\ref{fig:Q=3-5}, do not form linked or knotted loops.

\section{Conclusions}

The objective of this work is to extend further previous analysis of the solution of the $U(1)$ gauged Skyrme
model \cite{Piette:1997ny,Radu:2005jp,Livramento:2023keg} by considering the
solutions with topological charges up to $Q=5$
coupled to the magnetic field.
The gauged Skyrmions are topologically stable, they carry a local
magnetic flux which induces magnetic moment of the configuration.

We show that the the backreaction of the magnetic field strongly affects the usual structure of multisoliton
solutions, in particular the discrete  symmetry of ungauged Skyrmions becomes broken due to the presence of a circular magnetic flux.
The transformation to the toroidal Skyrmions is observed in the strong coupling regime for all
configurations we studied. Similar pattern was also observed for conventional Skyrmions in the model with symmetry breaking potential \cite{Gudnason:2014jga}.

The work here should be taken further by considering the electrically charged configurations with topological charge $Q>1$, the pattern of evolution of corresponding $Q=1$ gauged Skyrmion  was discussed in our previous work \cite{Livramento:2023keg}. It might be also interesting to consider gauged Skyrmions in a model with
a symmetry breaking potential.
We hope we can address these issues in our future work.

\section*{Acknowledgment}
We are grateful to E.~Radu and D.H.~Tchrakian for inspiring and valuable discussions. The numerical calculations were performed on the HybriLIT cluster at the JINR, Dubna.

%%%%%%%%%%%%%%%%%%%%%%%%%%%%%%%%%%%%%%%%%%%%%%%%%%%%%%%%%%%%%%%%%%
 \begin{small}
 
%%%%%%%%%%%%%%%%%%%%%%%%%%%%%%%%%%%%%%%%%%%%%%%%%%%%%%%%%%%%%%%%%%%%%%%%%%%%%%
 \end{small}
 \end{document}